\documentclass[preprint,5p,twocolumn,sort&compress]{elsarticle}

\usepackage{graphicx,epsfig}
\usepackage[english]{babel}
\usepackage{amssymb,amsfonts,amsmath}

  \newcommand{\cF}{{\cal F}}

  \newcommand{\cL}{{\cal L}}
\newcommand{\cM}{{\cal M}}  \newcommand{\cN}{{\cal N}}

\newcommand{\be}{\begin{equation}} \newcommand{\ee}{\end{equation}}
\newcommand{\bea}{\begin{eqnarray}} \newcommand{\eea}{\end{eqnarray}}
\newcommand{\beann}{\begin{eqnarray*}}  \newcommand{\eeann}{\end{eqnarray*}}
\newcommand{\bfig}{\begin{figure}} \newcommand{\efig}{\end{figure}}
\newcommand{\ba}{\begin{array}} \newcommand{\ea}{\end{array}}
\newcommand{\bcen}{\begin{center}} \newcommand{\ecen}{\end{center}}
\newcommand{\btab}{\begin{tabular}} \newcommand{\etab}{\end{tabular}}

\def\tr{\operatorname{tr\:}}

\newtheorem{Proposition}{Proposition}[section]

\newtheorem{Theorem}{Theorem}[section]
\newtheorem{Lemma}{Lemma}[section]
\newtheorem{Corrolary}{Corrolary}[section]

\newcommand{\bp}{\begin{Proposition}}   \newcommand{\ep}{\end{Proposition}}
\newcommand{\bt}{\begin{Theorem}}   \newcommand{\et}{\end{Theorem}}
\newcommand{\bl}{\begin{Lemma}}     \newcommand{\el}{\end{Lemma}}
\newcommand{\bc}{\begin{Corrolary}} \newcommand{\ec}{\end{Corrolary}}
\begin{document}


\title{Higher dimensional conformal field theories in the Coulomb branch}

\author[ch]{Carlos Hoyos}
\ead{choyos@phys.washington.edu}

\address[ch]{Department of Physics, University of Washington, Seattle, WA 98915-1560, USA}

\begin{abstract}

We use the AdS/CFT correspondence to study flows of $\cN=4$ SYM to non-conformal theories. The dual geometries can be seen as sourced by a Wigner's semicircle distribution of D3 branes. We consider two cases, the first case corresponds to a point in the Coulomb branch and the theory flows to a six dimensional conformal field theory. In the second case a mass is introduced for a hypermultiplet and the theory flows to a five dimensional conformal field theory. We argue from the gravity and the field theory side that the low energy theories correspond to the $(2,0)$ theory in six dimensions and to a theory with exceptional global symmetry $E_1$ in five dimensions.

\end{abstract}

\maketitle

 
\section{Introduction}

It has been known for some time how to describe generic points in the Coulomb moduli space of $\cN=4$ super Yang-Mills (SYM) theory using the AdS/CFT correspondence \cite{Maldacena:1997re,Witten:1998qj,Gubser:1998bc}. The expectation values of the six real scalar fields $\phi_i$  of $\cN=4$ SYM translate into a distribution of D3 branes in the transverse six dimensional space \cite{Kraus:1998hv}. In the near horizon limit the ten dimensional geometry takes the form
\begin{equation}
ds_{9,1}^2=H^{-1/2}\,\eta_{\mu\nu} dx^\mu dx^\nu+H^{1/2} \sum_{i=1}^6 (dy^i)^2,
\end{equation}
where the harmonic function $H(\vec{y})$ is determined by a distribution function $\sigma(\vec{w})$
\begin{equation}
H(\vec{y})=\int d^6 w\, \sigma(\vec{w} )\,\left|\vec{y}-\vec{w}\right|^{-4}.
\end{equation}
The geometry is asymptotically $AdS_5\times S^5$.

A special family of solutions was studied in ref. \cite{Freedman:1999gk} starting from $d=5$ $\cN=8$ 
supergravity (SUGRA) and uplifting to ten dimensions. The associated distribution functions $\sigma_n$ have support on a $n$-dimensional ball and preserve a $SO(n)\times SO(6-n)$ symmetry in the internal space. In the field theory, they correspond to expectation values of operators in the $\mathbf{20'}$ representation of the $SO(6)_R$ R-symmetry group of $\cN=4$ SYM, given by the symmetric traceless combinations $\tr \phi_{(i} \phi_{j)}$.
The most symmetric configuration, with $SO(5)$ symmetry, is a Wigner's semicircle distribution on an interval\footnote{The $n=5$ distribution is thought to be unphysical.}
\begin{equation}
\sigma_1(\vec{\phi})=\frac{2}{\pi \Lambda^2}\sqrt{\Lambda^2-\phi_1^2}\,\Theta(\Lambda^2-\phi_1^2)\prod_{i=2}^6\delta(\phi_i),
\end{equation}
associated to operators of the form $\sim \xi^{ij}\tr\left(\phi_i \phi_j\right)$, with $\xi^{ij}={\rm diag}\,(5,-1,-1,-1,-1,-1)$. This class of solutions has a gapless continuum spectrum, and at low temperatures the entropy density scales as $s\sim T^5$ \cite{Gubser:2000nd}.

In the $\cN=2^*$ SYM theory the field content of $\cN=4$ SYM is divided in an $\cN=2$ vector multiplet and a hypermultiplet, and a mass is introduced for the last.
The holographic duals for some configurations in the moduli space were constructed using a truncation of $d=5$ $\cN=8$ SUGRA \cite{Pilch:2000ue,Pilch:2003jg}. One class of solutions
also correspond to a Wigner's semicircle distribution 
\cite{Buchel:2000cn,Evans:2000ct,Carlisle:2003nd}. In this case the low temperature entropy density scales as $s\sim T^4$ \cite{Gubser:2000nd}. In addition, the speed of sound approaches the value $c_s^2=1/4$ and the bulk over shear viscosity ratio saturates the bound $\zeta/\eta\geq 2\left(1/3-c_s^2\right)=1/6$ \cite{Buchel:2007mf}. This is consistent with having a five dimensional conformal field theory (CFT) compactified on a circle \cite{Kanitscheider:2009as}. 

At finite temperature the Coulomb moduli space is lifted in general, but the low temperature regime is dominated by an infrared effective theory close to special points where the free energy is minimized (see e.g. \cite{Paik:2009iz}). In the cases we study, the local minima correspond to the semicircle distributions. We will show that the near horizon geometry of the zero temperature holographic duals can be seen as an $AdS_7$ compactified on a torus in the $\cN=4$ SYM case and an $AdS_6$ geometry compactified on a circle in the $\cN=2^*$ SYM case. Therefore, the dual effective theory is a $d=6$ CFT in the $\cN=4$ theory and a $d=5$ CFT in the $\cN=2^*$ theory, explaining the behavior of thermodynamic quantities. We will also argue that in the large-$N$ limit a sector of the $\cN=4$ low energy theory maps to the $(2,0)$ theory on the M5 brane \cite{Berkooz:1997cq} while in the $\cN=2^*$ theory a similar sector maps to a D4/D8/O8 intersection with $N_f=0$ flavors \cite{Brandhuber:1999np}.

Finally, we will argue that the mechanism giving rise to the effective higher dimensional theories is dimensional (de)construction \cite{ArkaniHamed:2001ca,ArkaniHamed:2001ie}. Given the distribution of eigenvalues in the Coulomb branch, the gauge group in these examples will be broken to a number of group factors of order $N$, and the spectrum of charged massive states will include masses scaling as $\sim 1/N$, that can be grouped to fill a Kaluza-Klein tower in the large-$N$ limit, effectively producing the additional dimensions.

\section{The $\cN=4$ flow to a $d=6$ CFT}

The holographic dual to the $SO(5)$ symmetric flow of $\cN=4$ SYM belongs to a larger family of solutions of maximally supersymmetric SUGRAs with non-trivial profiles for scalars belonging to the $SL(N,\mathbb{R})/SO(N)$ coset ($N=8,6,5$ for $d=4,5,7$ dimensions). The relevant action is \cite{Freedman:1999gk,Cvetic:1999xx}
\begin{equation}\label{eq:scalaraction}
e^{-1}\cL_d =R-\frac{1}{2}\sum_{i=1}^{N-1}(\partial\varphi_i)^2-V,
\end{equation} 
where the scalar potential is
\begin{equation}\label{eq:potscs}
V=-\frac{g^2}{2}\left[(\tr M)^2-2\tr M^2 \right],
\end{equation}
and $M$ is a diagonal $N\times N$ matrix with eigenvalues $X_i=e^{\beta_i}$, $i=1,\dots,N$ satisfying $\det M=1$. The exponents $\beta_i$ are linear combinations of the scalar fields $\varphi_i$, but we will not need the explicit expressions.

When $d=5$, the $SO(5)$ symmetric solution is
\begin{equation}
ds_{4,1}^2=(g r)^2 H^{1/6}\eta_{\mu\nu} dx^\mu dx^\nu+\frac{d r^2 }{(g r)^2 H^{1/3}}.
\end{equation}
Where $H=1+\ell^2/r^2$. The profile for the scalars $\beta_i=(1-6\delta_{i6})\beta/\sqrt{15}$ is determined by the flow equations for a single function $\beta$
\begin{equation}
H^{1/6} r\frac{d\beta}{dr}=-\frac{1}{2}\frac{\partial}{\partial \beta}\tr M.
\end{equation}
In the near horizon limit $u^2=1/(g^2\ell r)\to \infty$, the metric becomes
\begin{equation}\label{eq:n4metric}
ds_{4,1}^2\simeq \frac{(g\ell u)^{-4/3}}{u^2}\left[\eta_{\mu\nu} dx^\mu dx^\nu +\frac{4}{g^2} du^2 \right].
\end{equation}
The scalars asymptote to
\begin{equation}\label{eq:n4scalar}
X_{i<6}=X \simeq  2(g\ell u)^{2/3}, \ \ X_6 =X^{-5}\sim u^{-10/3} .
\end{equation}
In this limit the leading term in the scalar potential \eqref{eq:potscs} is
\begin{equation}\label{eq:n4potNH}
V_\infty = -\frac{15 g^2}{2} X^2.
\end{equation}
The metric \eqref{eq:n4metric} and scalar profiles \eqref{eq:n4scalar} are solutions of the action \eqref{eq:scalaraction} with the potential \eqref{eq:n4potNH}.

\subsection{Lift to eleven dimensions}

The maximally symmetric solution to \eqref{eq:scalaraction} in $d=7$ dimensions is an $AdS_7$ space
\begin{equation}\label{eq:7dmetric}
ds_{6,1}^2=\frac{1}{\tilde{u}^2}\left[\eta_{\mu\nu} d\tilde{x}^\mu d\tilde{x}^\nu+\delta_{ab} dy^a dy^b+\frac{4}{g^2} d\tilde{u}^2 \right].
\end{equation}
Here we set $y^a$, $a=1,2$ to be coordinates along a torus. The scalars have values $\tilde{X}_{i=1,\dots,5}=1$, so the value of the potential is
\begin{equation}\label{eq:potn4}
V_{d=7}=-\frac{15 g^2}{2}.
\end{equation}
Writing the metric \eqref{eq:7dmetric} as
\begin{equation}
ds_{6,1}^2=e^{-2\phi} ds_{4,1}^2+e^{3 \phi}(dy_1^2+dy_2^2),
\end{equation}
and reducing along the torus gives \eqref{eq:n4metric}, \eqref{eq:n4scalar} and \eqref{eq:n4potNH} if we identify
\begin{equation}
X=e^{-\phi}, \ \ \tilde{u}=2^{3/2} g\ell u, \ \ \ \tilde{x}^\mu=\sqrt{2} g\ell x^\mu.
\end{equation}

We have shown that the $\cN=4$ flow can be lifted to an $AdS_7$ solution of $d=7$ $\cN=2$ SUGRA. This geometry, in turn, can be lifted to an $AdS_7\times S^4$ solution of $d=11$ SUGRA, that is the near horizon geometry of a stack of M5 branes \cite{Cvetic:1999xx}. Therefore, in the large-$N$ and strong coupling approximation we are using, the infrared dynamics of the $SO(5)$ symmetric flow coincides with the $d=6$ $(2,0)$ CFT of the M5 brane at least in the subsector we have studied. In the field theory side this involves the components of the $d=6$ energy-momentum tensor $T_{\mu\nu}$, $T_{ab}$ that reduce to the $d=4$ energy-momentum tensor $T_{\mu\nu}$ plus a scalar field $T_{11}=T_{22}$, $T_{12}=0$. This is enough to explain the scaling of the entropy density $s\sim T^5$ and to make the prediction that, in the low temperature regime, the speed of sound will approach the value $c_s^2=1/5$, while the bulk over shear viscosity ratio will saturate the bound $\zeta/\eta\geq 2(1/3-c_s^2)=4/15$. 

\section{A large-$N$ equivalence for $\cN=2^*$ SYM}

An $\cN=2$ CFT that is equivalent to $\cN=4$ SYM in the large-$N$ limit can be constructed by doing a simple orientifold projection. The holographic dual is an $AdS_5\times S^5/\mathbf{Z}_2$ orbifold geometry, with an orientifold O7 plane and $N_f=4$ D7 branes sitting at the orbifold point  \cite{Fayyazuddin:1998fb}. The orientifold breaks the isometry group of the five-sphere $SO(6)\simeq SU(4)\to SU(2)\times SU(2)_R\times U(1)_R$ and does a $\mathbf{Z}_2\subset U(1)_R$ projection of supergravity fields. In the field theory this is interpreted in terms of the breaking of and the projection with respect to the R-symmetry group of the $\cN=4$ theory.
The field content is an $\cN=2$ theory with an $USp(2 N)$ vector multiplet, a hypermultiplet in the antisymmetric representation and $SO(8)$ flavor group. In terms of $\cN=1$ superfields the matter content is a vector multiplet $W_\alpha$, a chiral multiplet in the adjoint representation $X$, two chiral multiplets in the antisymmetric representation $A$, $\tilde{A}$ and 8 chiral multiplets in the fundamental representation $Q^i$, $\tilde{Q}_i$. There is a $USp(2)\simeq SU(2)$ symmetry that rotates the antisymmetric multiplets.  

The $\mathbf{20'}$ and $\mathbf{10}$ Kaluza-Klein modes of the dilaton and two-form potential are projected as
\begin{eqnarray}\label{eq:decomp}
\notag \mathbf{20'} & \to & (\mathbf{3},\mathbf{3})_0\oplus (\mathbf{1},\mathbf{1})_4\oplus (\mathbf{1},\mathbf{1})_{-4}\oplus (\mathbf{1},\mathbf{1})_0, \\
\mathbf{10} & \to & (\mathbf{3},\mathbf{1})_{-2}\oplus (\mathbf{1},\mathbf{3})_2.
\end{eqnarray}
The first row correspond to modes dual to operators of conformal dimension $\Delta=2$, while the second row is dual to $\Delta=3$ operators.

This can be compared with the $\cN=2^*$ theory. The breaking of supersymmetry implies that the maximal R-symmetry group is $SU(2)\times SU(2)_R\times U(1)_R$. Fields in the $\mathbf{20'}$ and $\mathbf{10}$ decompose similarly to \eqref{eq:decomp}, but in addition there are $(\mathbf{2},\mathbf{2})_{2}\oplus(\mathbf{2},\mathbf{2})_{-2}$ fields belonging to the $\mathbf{20'}$ representation and $(\mathbf{2},\mathbf{2})_{0}$ fields in the $\mathbf{10}$ representation. A mass term is added to the Lagrangian using the $(\mathbf{1},\mathbf{1})_0$ operator for scalars and the $(\mathbf{3},\mathbf{1})_2\oplus (\mathbf{3},\mathbf{1})_{-2}$ operator for fermions. 

Notice that the same modes appear in the orientifolded theory, so a mass term of this kind in the Lagrangian will give a mass to the antisymmetric chiral fields. In both cases, the R-symmetry group is reduced to $SU(2)_R\times U(1)$, with $U(1)\subset SU(2)$.

The large-$N$ equivalence of the massless theory implies an equivalence of the deformed theories as long as one is restricted to the common sector. When the $USp(2N)$ theory is at a generic point in the Coulomb branch of the moduli space, the couplings $Q^i X\tilde{Q}_i$ give a mass to the flavor fields. 

\subsection{$\cN=2^*$ flows as orientifold geometries}

The $\cN=2^*$ flows are solutions of the $d=5$ action \cite{Pilch:2000ue,Pilch:2003jg}
\begin{equation}\label{eq:action}
e^{-1}\cL_5= R-4(3\partial_\mu\alpha \partial^\mu \alpha+\partial_\mu\chi\partial^\mu \chi+ V(\alpha,\chi)).
\end{equation}
Where $R$ is the Ricci scalar. The scalar potential is
\begin{equation}\label{eq:scalpot}
\!\! V(\alpha,\chi)=-\frac{g^2}{ 4}\left[\frac{1}{\rho^4}+2\rho^2\cosh(2\chi)+\frac{1}{4}\rho^8 \sinh^2(2 \chi) \right],
\end{equation}
where $\rho=e^\alpha$. 

Supersymmetric solutions can be found solving a system of first order equations. 
Defining $c=\cosh(2 \chi)$, the result is
\begin{eqnarray}\label{eq:n2solutions}
\notag ds_{4,1}^2 & =&\frac{4}{g^2} \frac{d c^2}{\rho^8(c^2-1)^2}+k^2\frac{\rho^4}{c^2-1} \eta_{\mu\nu} d x^\mu dx ^\nu, \\
\rho^6 & = & c+(c^2-1)\left[\gamma+\frac{1}{2}\log\left(\frac{c-1}{c+1}\right) \right].
\end{eqnarray}
The boundary is at $c\to 1$, where the scalars vanish ($\rho=1$) and the metric asymptotes $AdS_5$.\footnote{There are several ways to lift these solutions to ten dimensions \cite{Pilch:2003jg}, but this will not be important for the analysis. In ref. \cite{Carlisle:2003nd} it was shown that some of the new solutions correspond to rotations of D3 brane distributions, this suggests that other solutions may correspond to different geometric transformations on the moduli space.} The flows are divided in three classes according to the value of the integration constant $\gamma$.
If $\gamma <0$, $\rho= e^\alpha$ vanishes for a finite value of $\chi$. If $\gamma =0$, $\rho\to 0$ as $\chi\to \infty$, this is the near horizon limit. Finally, if $\gamma >0$, both $\rho$ and $\chi$ diverge. The first class describes a distribution of D3 branes smeared on a circle around the origin of the moduli space, while the second class describes the Wigner's semicircle distribution. The third class is badly singular and will not be considered in the following.

The $d=5$ metric can be lifted to a solution of $d=10$ type IIB SUGRA. In the Einstein frame, the metric takes the form
\begin{equation}\label{eq:10dmetric}
ds_E^2=\Omega^2\rho^2 ds_{4,1}^2+\frac{4}{g^2}\Omega^2 ds_5^2,
\end{equation}
where the metric of the compact space is
\begin{equation}
ds_5^2=\frac{d \theta^2}{c}+\frac{\sin^2\theta}{X_2} d\phi^2+\rho^6 \cos^2\theta \left(\frac{\omega_3^2}{c X_2}+\frac{\omega_1^2+\omega_2^2}{X_1} \right).
\end{equation}
With
\begin{align}
\notag &X_1  =  \cos^2\theta+c \rho^6 \sin^2\theta, \ \
\notag &X_2  =  c \cos^2\theta+\rho^6 \sin^2\theta,\\
&\Omega^2  =  (c X_1 X_2)^{1/4}/\rho^3. \ \ &
\end{align}
and $\omega_i$ are the $SU(2)_R$ invariant forms parametrizing an $S^3$ of unit radius.
The geometry has background self-dual five-form flux $F_5=\cF+*\cF$, three-form flux $F_3=d A_2$ ($A_2=C_2+iB_2$),  and axion-dilaton fields $C_0+ie^{-\varphi}= i(1-B)/(1+B)$ given by
\begin{eqnarray}\label{eq:fluxes}
\notag \cF & = & dx^0 \wedge dx^1\wedge dx^2\wedge dx^3\wedge dw, \\
\notag A_2 & = & e^{i\phi} \left(i a_1 d\theta\wedge \omega_1+i a_2 \omega_2\wedge \omega_3+a_3 \omega_1 \wedge d\phi \right),\\
B & = & e^{2i\phi}(\sqrt{c X_1}-\sqrt{X_2})/(\sqrt{c X_1}+\sqrt{X_2}).
\end{eqnarray}
The values of the functions $w$, $a_1$, $a_2$ and $a_3$ can be found in original reference \cite{Pilch:2000ue}, they are real functions of $\theta$ and $\chi$ only and $a_i(\theta,\chi)=0$ when $\theta=\pi/2$ or $\chi=0$. Notice that the dilaton diverges at the locus $\theta=\pi/2$ when $c\to\infty$, so the theory becomes strongly coupled there.

Since this background geometry involves only fields common to the $\cN=2^*$ theory and the deformed $USp(2N)$ theory, it can be used as a holographic dual for both. The only difference is that for the $USp(2N)$ theory there is an orientifold O7 plane at $\theta=0$ that changes the periodicity of $\phi$ from $2\pi$ to $\pi$. Notice that the two-form is twisted, $A_2\to -A_2$ under a $\pi$ rotation of $\phi$, as is appropriate for an O7 plane.

The non-zero dilaton flux with a trivial monodromy around the $\phi$ direction can be interpreted as produced by D7 brane charge density in the orientifold geometry. Since $B\to 0$ as $c\to 1$, the D7 brane charge density vanish at the $AdS_5$ boundary, so there are $N_f=4$ D7 branes at the orientifold point $\theta=0$ that compensate the flux. In the field theory the flavor fields acquire a mass through the expectation value of the adjoint scalar, thus in the dual the D7 branes move away from the orientifold point inducing a charge density distribution in the bulk. 

\section{The $\cN=2^*$ flow to a $d=5$ CFT}

We have shown the large-$N$ equivalence of the $\cN=2^*$ theory and the massive $USp(2N)$ theory, so we will be able to work mostly with the more intuitive orientifold interpretation of the holographic dual in the following. In order to describe the infrared physics we will now study the near horizon limit from both the five and the ten dimensional points of view.

\subsection{Near horizon limit in five dimensions.}\label{sec:pwflow}

The near horizon limit for the second class of solutions in \eqref{eq:n2solutions} ($\gamma=0$) correspond to $\chi\to\infty$. Introducing a new coordinate $u\to \infty$, we have
\begin{equation}\label{eq:nhsols}
e^{2\chi}\simeq 2 u, \ \ e^{6 \alpha}\simeq 2/(3 u), \ \ e^A\simeq 2^{1/3}k u^{-4/3}/3^{1/3}.
\end{equation}
The metric in this limit is
\begin{equation}\label{eq:5dmetric}
ds_{4,1}^2\simeq \left(\frac{3}{2} \right)^{4/3} u^{-8/3}\left[\frac{4}{g^2} du^2+\left(\frac{ 2k}{3}\right)^2 \eta_{\mu\nu} dx^\mu dx^\nu\right].
\end{equation}
One can find these expressions as solutions of the action \eqref{eq:action} with a modified potential that keeps only the leading terms of the $\chi\to \infty$ limit. Defining
\begin{equation}\label{eq:scalars}
\phi_1=\frac{1}{2}(3\alpha+\chi), \ \ \phi_2=\frac{1}{2}(\alpha-\chi),
\end{equation}
the potential \eqref{eq:scalpot} factorizes in this limit $V_\infty(\phi_1,\phi_2)   =   e^{-2\phi_2} V_1(\phi_1)$, where
\begin{equation}\label{eq:pot2}
V_1(\phi_1)  =  -\frac{g^2}{4}\left[e^{-2\phi_1}+e^{2\phi_1}-\frac{1}{16} e^{6 \phi_1}  \right]. 
\end{equation}

\subsection{Lift to six dimensions.}\label{sec:uplift}

We now proceed to lift the near horizon geometry to six dimensions. It turns out that the geometry is a solution to $d=6$ $\cN=(1,1)$ $F(4)$ SUGRA  \cite{Romans:1985tw}. This result can be obtained starting with the $d=6$ action
\begin{equation}\label{eq:6daction}
e^{-1}\cL_6= R-4(\partial_\mu\phi_1 \partial^\mu \phi_1+V_1(\phi_1)),
\end{equation}
where the potential $V_1(\phi_1)$ has been defined in \eqref{eq:pot2}. One direction, $x_6$, is taken to be compact. The ansatz for the metric is 
\begin{equation}\label{eq:6dmetric}
ds_{5,1}^2=e^{-2\phi_2} ds_{4,1}^2+e^{6\phi_2} dx_6^2,
\end{equation}
where $ds_{4,1}^2$ is the five-dimensional metric \eqref{eq:5dmetric}. Reducing along the $x_6$ direction and using \eqref{eq:scalars}, one recovers the $d=5$ action \eqref{eq:action} with the near horizon potential $V_\infty(\phi_1,\phi_2)$. Now, from \eqref{eq:nhsols} and \eqref{eq:scalars} the value of the scalars in the near horizon solution is
\begin{equation}\label{eq:vevscalars}
e^{4\phi_2}= u^{-4/3}/(12)^{1/3}, \ \ e^{4\phi_1}=4/3.
\end{equation}
Using this and \eqref{eq:5dmetric} in \eqref{eq:6dmetric}, the lift of the metric is locally $AdS_6$ space
\begin{equation}
ds_{5,1}^2=\frac{3^{3/2}}{2 u^2}\left[ \frac{4}{g^2} du^2 + \left(\frac{2 k}{3} \right)^2\eta_{\mu\nu} dx^\mu dx^\nu+\frac{1}{9} d x_6^2\right].
\end{equation}
Now it is possible to see that this solution matches with the maximally supersymmetric solution of $F(4)$ SUGRA.  

The Einstein plus scalar action of $F(4)$ SUGRA is of the form \eqref{eq:6daction} with a potential that depends on two parameters, the $d=6$ gauge coupling $g_6$ and the two-form ``mass'' $m$ \cite{Romans:1985tw} 
\begin{equation}\label{eq:f4potential}
V_{d=6}(\phi)=-\frac{1}{8}\left[g_6^2 e^{2 \phi}+4 g_6 m e^{-2 \phi}-m^2 e^{-6 \phi}\right].
\end{equation}
The scalar field $e^\phi$ is normalized\footnote{The field $\phi$ used here is $\sqrt{2}$ times the one used in ref. \cite{Romans:1985tw}, in order to match with the kinetic coefficients in \eqref{eq:6daction}. This does not affect to the discussion.} in such a way that the maximally supersymmetric $g_6=3m$ and nonsupersymmetric $g_6=m$ critical points of the potential appear at $e^\phi=1$. At these points the geometry is $AdS_6$, so the normalization of $e^{\phi_1}$ should match with the one of $e^\phi$. Comparing with \eqref{eq:vevscalars} and the sign of the exponentials in \eqref{eq:pot2} and \eqref{eq:f4potential}, this means the field $\phi_1$ should be $\phi_1=-\phi+\log(4/3)/4$. Then, the potential \eqref{eq:pot2} 
coincides with \eqref{eq:f4potential} when $g_6= 3m$ and $g_6^2=\sqrt{3} g^2$.\footnote{The degrees of freedom of the $d=5$ $\cN=2$ SUGRA we are considering match with the dimensional reduction of $\cN=(1,1)$ $d=6$ $F(4)$ SUGRA with and additional $U(1)$ vector multiplet, constructed in refs. \cite{D'Auria:2000ad,Andrianopoli:2001rs}. The scalar manifolds also coincide, although the parametrization seems to be different.}

\subsection{Massive T-duality in ten dimensions.}

The $AdS_6$ factor can also be observed in the ten dimensional geometry \eqref{eq:10dmetric} using a generalization of T-duality. Both the dilaton and the potential $A_2$ in \eqref{eq:fluxes} have a simple phase dependence on the angular coordinate $\phi$. This can be seen as a $SO(2)\subset SU(1,1)\cong SL(2,\mathbb{R})$ fibration along the compact $\phi$ direction, and in the orientifold geometry this has a simple interpretation in terms of D7 branes. These kind of backgrounds were studied in ref. \cite{Meessen:1998qm} and a set of generalized T-duality rules was derived there, one simply applies the usual rules  
but dropping the $\phi$ dependence. However, the T-dual theory is not type IIA anymore, but a massive generalization of it with a potential for the axion-dilaton fields.

The dilaton and the two-form potential written in $SL(2,\mathbb{R})$ covariant form are
\begin{equation}
\cM=e^{\varphi}\left(\begin{array}{cc} C_0^2+e^{-2\varphi} & C_0 \\ C_0 & 1 \end{array}\right), \ \ \vec{B}_2=\left(\begin{array}{c} C_2 \\ B_2 \end{array} \right).
\end{equation}
A general $SL(2,\mathbb{R})$ fibration has the form of a local transformation $\Lambda(\phi)$
\begin{equation}
\cM(\phi) = \Lambda(\phi) \cM^b \Lambda(\phi)^T, \ \ \vec{B}_{2}(\phi)=\Lambda(\phi)\vec{B}_2^b,
\end{equation}
where the bare values $\cM^b$ and $\vec{B}_2^b$ do not depend on $\phi$. The transformation can be parametrized as
\begin{equation}
\Lambda(\phi)=e^{\frac{1}{2}\phi m^i T_i}=e^{\phi \tilde{m}}, \ \ T_1=\sigma_3, \ T_2=\sigma_1, \ T_3=i\sigma_2,
\end{equation}
where $\sigma_i$, $i=1,2,3$ are the Pauli matrices. An $SO(2)$ fibration has parameters $m^1=m^2=0$, $m^3=m$. In this case, the axion-dilaton potential in the T-dual theory is
\begin{multline}
V(\cM)=\frac{1}{2}{\rm Tr}\,\left( \tilde{m}^2+\tilde{m}\cM\tilde{m}^T\cM^{-1}\right) = \\ = \frac{m^2}{8}e^{2 \varphi }C_0^2 (C_0^2+2) +\frac{m^2}{4}\cosh(2\varphi) +\frac{m^2}{4} (C_0^2-1).
\end{multline}

The bare values of the fields that will be needed to compute the T-dual axion, dilaton and metric are the axion, dilaton and NS two-form
\begin{equation}
\notag C_0^b   =  0, \ \
\notag e^{-\varphi^b}  = (X_2/c X_1)^{1/2},\ \ 
B_{\mu\phi}^b  =  0,
\end{equation}
and the Einstein metric $G_{MN}$ that can be read directly from \eqref{eq:10dmetric}. We will perform the T-duality on the strongly coupled region, the near horizon limit close to the $\theta=\pi/2$ locus.\footnote{Close to the orientifold the T-dual metric is singular, this might be solved by an uplift to a massive version of $d=11$ SUGRA (e.g.~\cite{Meessen:1998qm}).}

Expressions simplify in this case, it will be enough to check the scaling with the radial coordinate. For the type IIB metric
\begin{multline}
G_{\phi\phi} \sim u^{3/2}, \ \  G_{ab}\sim u^{-1/2}, \ \  G_{\mu\nu}\sim u^{-5/2},\\  G_{uu} \sim u^{-5/2}, \ \  e^{\varphi^b} \sim u. 
\end{multline}
With $\mu,\nu$ the spacetime indexes and $a,b$ the indexes along the three-sphere.

Applying T-duality rules, the dual axion vanishes $C_0=0$ and the dual dilaton and metric in the string frame scale as
\begin{align}
\notag  &\widetilde G_{\phi\phi}^s=e^{-\varphi^b/2} G_{\phi\phi}^{-1} \sim u^{-2}, &\widetilde G_{ab}^s  = e^{\varphi^b/2} G_{ab} \sim 1.  \\
 \notag &\widetilde G_{\mu\nu}^s  = e^{\varphi^b/2} G_{\mu\nu} \sim u^{-2}, &\widetilde G_{uu}^s  = e^{\varphi^b/2} G_{uu} \sim u^{-2}, \\
 &e^{\varphi} = e^{3\varphi^b/4}G_{\phi\phi}^{-1/2}\sim 1, \ 
\end{align}
Therefore the T-dual of the near-horizon region describes a locally (in the internal space) $AdS_6\times \mathbb{R}^4$ geometry, with $\phi$ a compact direction along the $AdS_6$. Since the axion vanishes, the potential for the dilaton is simplified to
\begin{equation}\label{eq:dilatpot}
V(\varphi)=\frac{m^2}{2} \sinh^2\varphi,
\end{equation}
and in this case it is just a constant. Summarizing, from the point of view of massive type IIA theory the singular region of type IIB is just an $AdS_6$ space with a constant dilaton. Interestingly, the reduction of a warped product of $AdS_6\times S^4$ solution of Roman's massive type IIA theory \cite{Romans:1985tz} on the $S^4$ gives an $AdS_6$ solution of $F(4)$ SUGRA in six dimensions \cite{Cvetic:1999un}. It would be interesting to see if this version of massive type IIA has similar solutions and reduces to the geometry we have found previously.

\subsection{Low energy effective theory}

Since the near-horizon geometry is an $AdS_6$ space, the infrared physics of the dual theory should be a $d=5$ CFT. We have obtained this result through a T-duality on a background containing O7 planes and D7 branes, so a candidate is the CFT living at D4/D8/O8 intersections \cite{Brandhuber:1999np}. This is confirmed by the lift to $d=6$ $F(4)$ SUGRA, that can in turn be lifted to the D4/D8/O8 geometry \cite{Cvetic:1999un}. We can also see that this picture is consistent with the T-dual geometry.

In the configurations of interest D4 branes intersect an O8 plane localized at an orbifold point\footnote{Therefore, for $N$ D4 branes the gauge group is $Sp(2N)$.}, and there are 8 D8-branes that compensate the charge of the orientifold planes. When a D8 brane coincides with the stack of D4 branes, a hypermultiplet in the fundamental representation of the gauge group becomes massless. In principle the global symmetry group of the gauge theory on the D4 branes includes an $SO(2 N_f)$ flavor symmetry, where $N_f$ is the number of massless hypermultiplets, a $U(1)_I$ symmetry associated to instanton number conservation and an $SU(2)$ symmetry due to the massless antisymmetric hypermultiplets. At the origin of the moduli space the global symmetry $SO(2 N_f)\times U(1)_I$ is enhanced to $E_{N_f+1}$\footnote{$E_{N_f+1}=$ $E_8$, $E_7$, $E_6$, $E_5={\rm Spin}(10)$, $E_4=SU(5)$, $E_3=SU(3)\times SU(2)$, $E_2=SU(2)\times U(1)$, $E_1=SU(2)$.} and the theory is conformally invariant \cite{Seiberg:1996bd}.

In the case at hand the Coulomb moduli space of the $d=5$ theory can be explored introducing a probe D4 brane in the type IIA geometry, extending along the $x^\mu$ and $\phi$ directions.
If the D4 brane is localized close to $\theta=\pi/2$, the isometries of the internal space are enhanced to $SO(4)\simeq SU(2)\times SU(2)$ in the near horizon limit, recovering the symmetries of the D4/D8/O8 intersection. This is possible if the near horizon geometry is describing a special point on the Coulomb moduli space that in the large-$N$ limit corresponds to a conformal theory with $SU(2)\times E_1$ ($N_f=0$) global symmetry.
The equivalence between the infrared theory of the $\cN=2^*$ flow and the $d=5$ $E_1$ CFT is expected to hold at least in the scalar sector we have studied, this is enough to explain the observed thermodynamic properties and should fix other quantities like the two-point functions of scalar operators.

\section{Deconstruction in the Coulomb branch}

So far we have determined the low energy effective theories associated to the near horizon geometry using SUGRA arguments, we will present now a field theory analysis to explain the appearance of higher dimensional theories at low energies, using arguments similar to those employed for the deconstruction of the $(2,0)$ six-dimensional theory in ref. \cite{ArkaniHamed:2001ie}. 

In the first example one of the six real scalars of the $\cN=4$ theory acquires an expectation value
\begin{equation}
\left\langle\mathbf{\phi} \right\rangle=\left(\begin{array}{cccc} \varphi_1 & & & \\ & \varphi_2 & & \\ & &  \ldots & \\ & & &  \varphi_N  \end{array} \right)\,,
\end{equation}
where the eigenvalues $\varphi_i$ are distributed on an interval $-\Lambda<\varphi_i<\Lambda$. The distribution is the Wigner's semicircle distribution, that on the real line takes the form
\begin{equation}
\rho(\varphi)=\frac{2 N}{\pi \Lambda^2}\sqrt{\Lambda^2-\varphi^2}.
\end{equation}
Consider the following change of variables
\begin{equation}
\varphi=\frac{\Lambda\hat{\varphi}}{N}.
\end{equation}
The new dimensionless variable is defined in the interval $\hat{\varphi}\in(-Nc,N)$. The eigenvalue distribution becomes
\begin{equation}
\hat{\rho}(\hat{\varphi})=\frac{2}{\pi}\sqrt{1-\frac{\hat{\varphi}^2}{N^2}}.
\end{equation}
In the large-$N$ limit the distribution becomes a constant $\hat{\rho}_\infty(\hat{\phi})=\frac{2}{\pi}$ defined on the real line $\hat{\varphi}\in(-\infty,\infty)$. The distribution is such that there is one eigenvalue in an interval of length $\Delta \hat{\varphi}=\pi/2$. This limit simply focus on the region of moduli space close to the origin $\varphi=0$. Other limits are possible if one focus around a different region, for instance if one defines 
\begin{equation}
\varphi=\Lambda \eta+\frac{\Lambda\hat{\varphi}}{N},
\end{equation}
with $-1<\eta<1$ and $\hat{\varphi}\in (-N(1+\eta),N(1-\eta))$. Provided $\sqrt{1-\eta^2}>O(1/N)$, in the large-$N$ limit the distribution extends to the real line and becomes a constant $\hat{\rho}_\infty(\hat{\phi})=\frac{2}{\pi}\sqrt{1-\eta^2}$.

If we zoom on a region around the origin of moduli space, the size of the interval with just one eigenvalue in the original variable is
\begin{equation}
\Delta\varphi = \frac{\Lambda}{2} \left(\frac{\pi}{N}\right).
\end{equation}
This is also the separation between eigenvalues, so the low energy effective theory reduces to a $U(1)^{N-1}$ gauge theory with a massive tower of 1/2 BPS vector multiplets charged under different $U(1)$ groups. The lowest states have a mass 
\begin{equation}
m_n = \frac{g_{YM} \Lambda}{2} \left(\frac{\pi}{N}\right) n, \ \ n=1,2,3,\dots.
\end{equation}
This is a good approximation up to values of masses that are sensible to the deviation from an uniform distribution, which will be a fraction of $\Lambda$ but $O(N)$ times the masses of the lowest states. Then, the low energy states resemble a tower of Kaluza-Klein (KK) modes with momentum along a circle of length
\begin{equation}
L_5 = \frac{ 4 N}{g_{YM} \Lambda} \sim N^{3/2} (\lambda_{YM})^{-1/2}\Lambda^{-1},
\end{equation}
where $\lambda_{YM}=g_{YM}^2 N$ is the 't Hooft coupling. In addition to this set of states, the $SL(2,\mathbb{Z})$ S-duality of the $\cN=4$ theory implies that there are also towers of dyonic states, including magnetically charged states with masses
\begin{equation}
M_n \simeq \frac{\Lambda}{2 g_{YM}} \left(\frac{\pi}{N}\right) n, \ \ n=1,2,3,\dots.
\end{equation}
These can be mapped to KK modes with momentum along a different circle of length
\begin{equation}
L_6 = \frac{ 4g_{YM} N}{\Lambda}\sim N^{1/2} (\lambda_{YM})^{1/2} \Lambda^{-1}.
\end{equation}
Dyonic states map to KK modes with momentum along both circles. With this interpretation, the six-dimensional gauge coupling should be
\begin{equation}
g_6^2=g_{YM}^2 L_5 L_6 =\frac{16 g_{YM}^2 N^2}{\Lambda^2}\sim  N \lambda_{YM} \Lambda^{-2}.
\end{equation}
So in the large-$N$ limit there is a hierarchy of scales $L_5\gg L_6 \sim g_6 \gg \Lambda^{-1} $. Notice that there are some ambiguities in the way we construct the higher dimensional theory from the distribution of eigenvalues. We could have adopted the point of view that the gauge group is broken to $U(k)^{(N-1)/k}$, where $k$ is a divisor of $N-1$. This corresponds to having stacks of $k$ coincident eigenvalues, separated by an interval $k\Delta\varphi$. The spectrum and the length of the compact directions will change simply as $m_n\to k m_n$, $M_n\to k M_n$, $L_5\to L_5/k$, $L_6\to L_6/k$, $g_6\to g_6/k$. Clearly, for $k\ll N$, this does not affect to the argument. Also, we could have zoomed on a different region on moduli space, the only difference will be again a change in the size of the separation between eigenvalues $\Delta\varphi\to \Delta\varphi/\sqrt{1-\eta^2}$, producing the corresponding rescaling of the spectrum.

Summarizing, in the large-$N$ limit the low energy effective theory maps to a six-dimensional theory compactified on a torus. Since it corresponds to a point in the Coulomb branch of $\cN=4$ SYM, the six-dimensional theory will have 16 supercharges, and the BPS spectrum coincides with that of the $(2,0)_k$ theory \cite{ArkaniHamed:2001ie}. The limits $N\to\infty$, $g_{YM}^2 N\gg 1$ that are taken in the holographic dual correspond precisely to this situation. Notice that in this effective six-dimensional theory the compactification radii are also very large compared to the cutoff scale $\Lambda^{-1}$ where deviations from the six-dimensional behavior should be observed. This give us an explanation from the field theory perspective of the near horizon $AdS_7$ geometry we observe in the holographic dual.

We can apply the same arguments to the $\cN=2^*$ SYM theory at the corresponding point in the Coulomb branch, but in this case the $SL(2,\mathbb{Z})$ invariance is lost, so there is no tower of dyonic modes dual to the multiplets of the $W$ bosons. Therefore the deconstructed theory will have only one compact direction. Notice also that the hypermultiplet has a mass $\sim \Lambda$, so it decouples from the low energy theory. This leave us with an effective five-dimensional gauge theory with 8 supercharges and $U(k)$ gauge group. From the holographic calculation we know it is conformal and there are no flavors, so it should correspond to the point in moduli space with enhanced $E_1$ symmetry. Notice that, in the limit where the gravity dual geometry is a valid description, the five-dimensional coupling diverges 
\begin{equation}
g_5^2=g_{YM}^2 L_5 =\frac{4 g_{M} N}{k m}\to \infty.
\end{equation}
If we interpret $g_5$ as the bare coupling of the theory this corresponds precisely to where the conformal fixed point is expected to be \cite{Seiberg:1996bd}, so also in this case there is a consistent interpretation of the near horizon $AdS_6$ geometry from the field theory side.

\section*{Acknowledgments}
I am indebted to Andreas Karch for many useful comments and suggestions. I also want to thank Lawrence G. Yaffe for useful comments. I thank the Galileo Galilei Institute for Theoretical Physics for the hospitality and the INFN for partial support during the completion of this work. This work was supported in part by the U.S. Department of Energy under grant DE-FG02-96ER40956.

\bibliographystyle{elsarticle-num}
\bibliography{N2theoryB}


\end{document}